\documentstyle[12pt]{article}
\begin{document}
\newcommand{\cl}{\centerline}
\renewcommand{\theequation}{\arabic{equation}}
\newcommand{\beq}{\begin{equation}}
\newcommand{\eeq}{\end{equation}}
\newcommand{\bea}{\begin{eqnarray}}
\newcommand{\eea}{\end{eqnarray}}
\newcommand{\nn}{\nonumber}
\newcommand\pa{\partial}
\newcommand\un{\underline}
\newcommand\ti{\tilde}
\newcommand\pr{\prime}
\begin{titlepage}
\setlength{\textwidth}{5.0in} \setlength{\textheight}{7.5in}
\setlength{\parskip}{0.0in} \setlength{\baselineskip}{18.2pt}

\hfill {\tt SOGANG-HEP 302/03}
\begin{center}
{\large\bf Complete BFT Embedding of Massive Theory with One- and
Two-form Gauge Fields}
\end{center}
\vskip 0.5cm

\begin{center}
{Seung-Kook Kim$^{}$\footnote{\small Electronic address:
skandjh@empal.com}, Yong-Wan Kim$^{}$\footnote{\small Electronic
address: ywkim65@netian.com} and Young-Jai
Park$^{}$\footnote{\small Electronic address:
yjpark@ccs.sogang.ac.kr}}
\end{center}

\begin{center}
{$^{1}$ Department of Physics, Seonam University, \\
Namwon, Chonbuk 590-170, Korea\\
$^{2}$ Department of Physics and Institute for Science and
Technology,\\
Sejong University, Seoul 143-747, Korea\\
$^{3}$Department of Physics and Basic Science Research Institute,\\
Sogang University, C.P.O. Box 1142, Seoul 100-611, Korea\\}
\end{center}

\begin{center}
(\today)
\end{center}

\vfill
\begin{center}
{\bf ABSTRACT}
\end{center}
\begin{quotation}
We study the constraint structure of the topologically massive
theory with one- and two-form fields in the framework of
Batalin-Fradkin-Tyutin embedding procedure. Through this analysis
we obtain a new type of Wess-Jumino action with novel symmetry,
which is originated from the topological coupling term, as well as
the St\"uckelberg action related to the explicit gauge breaking
mass terms from the original theory.

\vskip 0.5cm

\end{quotation}
\end{titlepage}

\newpage
\section{Introduction}

Dirac scheme in the Hamiltonian formalism \cite{dirac} has been
widely used to quantize second-class constraint system in which
the Poisson brackets of constraints do not vanish on constraint
surface. For second-class constraint system, however, there are
under unfavorable circumstances in finding canonically conjugate
pairs since the resulting Dirac brackets may be in general
field-dependent and/or nonlocal, and have a serious ordering
problem between field operators. On the other hand, the
quantization of first-class constraint system
\cite{fradvil75,henneaux85} has been well appreciated in a gauge
invariant manner preserving Becci-Rouet-Stora-Tyutin symmetry
\cite{brst}. Therefore, if second-class constraint system can be
converted into first-class one using auxiliary degrees of freedom
to extend a phase space, we do not actually need to define Dirac
brackets and then the remaining quantization program follows the
method of Ref. \cite{fradvil75,henneaux85,brst}. This procedure
has been established by Batalin, Fradkin, and Tyutin (BFT)
\cite{bf86,bt91} and extensively studied in the canonical
formalism for various models including massive gauge fields
\cite{proca}, spontaneously broken gauge theories \cite{sbgt},
$SU(2)$/$SU(3)$ Skyrmion models \cite{skyrme}, non-linear sigma
models \cite{sigma}, noncommutative systems \cite{noncom}, and
many others \cite{other}. The BFT embedding formalism has been
also widely applied to obtain the Wess-Zumino (WZ) actions
\cite{wz} showing that original theory can be regarded as a
gauge-fixed version of extended gauge system, while verifying dual
equivalent descriptions \cite{bft-dual} in particular gauges from
the phase space partition function corresponding to the BFT
embedded involutive Hamiltonian.

On the other hand, antisymmetric tensor fields appearing first as
a mediator of the interaction \cite{kalb} have been much
interested in as an alternative of the Higgs mechanism without
residual Higgs scalar \cite{djt,alt-higgs}. With the topologically
interacting terms of the form $B\wedge F$, this mechanism is
considered generic in string phenomenology \cite{string}.
Moreover, various dual descriptions between different models have
been widely studied where antisymmetric tensor fields play an
important role in realization of dualities \cite{dual,duff}. In
particular, several years ago Banerjee and Banerjee had studied a
master Lagrangian \cite{banerjee} which is a first order massive
spin-one theory involving antisymmetric tensor fields in order to
show the dual equivalence of the Proca model and the massive
Kalb-Ramond model within a path integral framework. Recently,
Harikumar and Sivakumar \cite{harikumar} have elaborated on the
Lagrangian through the Hamiltonian and Lagrangian embedding
technique, and have shown that the embedded theory with
appropriate gauge fixing is equivalent to the $B\wedge F$ theory
on the level of Hamiltonian. The Lagrangian regarded as a
topologically massive theory has interesting constraint structure
related to the topological coupling term as well as the gauge
symmetry breaking mass term. As results, the second-class
constraints appears from two different origins: One comes from the
explicit gauge symmetry breaking mass term, and the other from the
topological coupling term. Since this theory have been studied by
a reduction of the set of constraints through Faddeev-Jackiw
method \cite{fj}, there still remains to be clarified the role of
auxiliary fields, which are originated from the topological
interaction term.

In the present paper, we shall fully apply the BFT method to the
topologically massive theory with one- and two-form gauge fields.
As results, we will explicitly show that auxiliary fields in part
for the BFT embedding are nothing but the well-known St\"uckelberg
fields on one hand, and find a new type of WZ action \cite{nwz}
composed of the remaining auxiliary fields, which are originated
from symplectic structure of the model on the other hand. We also
clarify the underlying constraint structure concerning irreducible
and reducible constraints.

In section 2, we newly construct the constraint structure of this
model, which is much simpler than that of the previous work
\cite{harikumar} due to the absence of derivatives in their
Poisson brackets. In section 3, we carry out the complete BFT
embedding of the theory which has local gauge symmetry
supplemented with other: Local gauge symmetry is recovered from
explicitly gauge symmetry breaking mass terms and the other comes
from underlying symplectic structure of topological term. In
section 4, by identifying new auxiliary degrees of freedom with
St\"uckelberg vector fields and new type of WZ fields, we obtain
simultaneously the St\"uckelberg Lagrangian related to the
explicit gauge breaking mass terms and a new type of WZ action
with novel symmetry, which is originated from the symplectic
structure of the theory. In section 5, we revisit the original
theory making use of gauging technique to show the equivalence of
the gauged Lagrangian and the St\"uckelberg Lagrangian on the
constraint surface. Conclusion is devoted in section 6.

\section{Constraint Structure of Topologically Massive Theory}
\setcounter{equation}{0}
\renewcommand{\theequation}{\arabic{section}.\arabic{equation}}

In this section, we consider topologically massive theory with
one- and two-form fields, described by the following first order
Lagrangian \cite{banerjee,harikumar}:
\begin{equation}
\label{Lag} {\cal L}=
-\frac{1}{4}B_{\mu\nu}B^{\mu\nu}+\frac{1}{2}A_\mu A^\mu
+\frac{1}{2m}\epsilon_{\mu\nu\rho\sigma}B^{\mu\nu}\partial^\rho
A^\sigma.
\end{equation}
From the symmetrized form of the Lagrangian of
\begin{equation}
\label{sym-Lag} {\cal L}=
-\frac{1}{4}B_{\mu\nu}B^{\mu\nu}+\frac{1}{2}A_\mu A^\mu
+\frac{1}{4m}\epsilon_{\mu\nu\rho\sigma}B^{\mu\nu}\partial^\rho
A^\sigma -\frac{1}{4m}\epsilon_{\mu\nu\rho\sigma}\partial^\mu
B^{\nu\rho} A^\sigma,
\end{equation}
we read the canonical momenta as
\begin{eqnarray}
\label{momenta}
&& \pi_0 = 0, ~~~\pi_i = \frac{1}{4m}\epsilon_{ijk}B^{jk}, \nonumber\\
&& \pi_{0i} = 0, ~~~\pi_{ij} = - \frac{1}{2m}\epsilon_{ijk}A^k,
\end{eqnarray}
where we denote $\epsilon_{0ijk}=\epsilon_{ijk}$ and
$\epsilon^{123}=+1$. Then, the primary Hamiltonian yields
\begin{equation}
\label{pri-H} {\cal H}_p={\cal H}_c + \lambda^0 \pi_0 + \lambda^i
\Omega_i +\Sigma^{0i}\pi_{0i}+\Sigma^{ij}\Omega_{ij}
\end{equation}
with the Lagrange multipliers $\lambda^0$, $\lambda^i$,
$\Sigma^{0i}$, and $\Sigma^{ij}$, where the canonical Hamiltonian
is given by
\begin{eqnarray}
\label{can-H} {\cal
H}_c&=&\frac{1}{4}B_{ij}B^{ij}-\frac{1}{2}A_iA^i
+\frac{1}{2}B_{0i}B^{0i}-\frac{1}{2}A_0A^0\nonumber\\
&-&\frac{1}{m}\epsilon^{ijk}B_{0i}\partial_j
A_k-\frac{1}{2m}A^0\epsilon^{ijk}\partial_iB_{jk},
\end{eqnarray}
and the primary constraints are defined as
\begin{eqnarray}
\label{pri-con} && \pi_0\approx 0,
~~~\Omega_i\equiv\pi_i-\frac{1}{4m}\epsilon_{ijk}B^{jk}\approx 0,\nonumber\\
&& \pi_{0i}\approx 0,~~~\Omega_{ij}\equiv\pi_{ij}+
\frac{1}{2m}\epsilon_{ijk}A^k\approx 0.
\end{eqnarray}
From the time stability conditions of the constraints $\pi_0$ and
$\pi_{0i}$, we have obtained two additional secondary constraints
as
\begin{eqnarray}
\label{sec-con} \Lambda&\equiv&\dot{\pi}_0=\{\pi_0, {\cal
H}_p\}=A_0+\frac{1}{2m}\epsilon^{ijk}\partial_iB_{jk}\approx
0,\nonumber\\
\Lambda_i&\equiv&\dot{\pi}_{0i}=\{\pi_{0i}, {\cal H}_p
\}=-B_{0i}+\frac{1}{m}\epsilon_{ijk}\partial^j A^k\approx 0,
\end{eqnarray}
and the constraints $\Omega_i$, $\Omega_{ij}$ fix the Lagrange
multipliers $\Sigma^{ij}$, $\lambda^i$, respectively. The other
Lagrange multipliers $\lambda^0$, $\Sigma^{0i}$ are also
determined by requiring consistency of the secondary constraints
$\Lambda$, $\Lambda_i$ with the equations of motion, and thus no
further new constraints are generated. As a result, the Poisson
brackets of all the constraints (\ref{pri-con}) and
(\ref{sec-con}) are obtained as
\begin{eqnarray}
\label{pb}
\{\pi_0, \Lambda\}&=& -\delta(x-y),\nonumber\\
\{\pi_{0i}, \Lambda_j \}&=& \delta_{ij}\delta(x-y),\nonumber\\
\{\Omega_i, \Omega_{jk} \}&=& -\frac{1}{m}\epsilon_{ijk}\delta(x-y),\nonumber\\
\{\Omega_i, \Lambda_j \}&=& \frac{1}{m}\epsilon_{ijk}\partial^k_x\delta(x-y),\nonumber\\
\{\Omega_{ij},\Lambda\}&=&\frac{1}{m}\epsilon_{ijk}\partial^k_x\delta(x-y)
\end{eqnarray}
showing that the constraint structure of the topologically massive
theory be fully second-class.

Now, instead of reducing the constraints $\Omega_i$, $\Omega_{ij}$
strongly by making use of Faddeev-Jackiw scheme \cite{fj} and
considering only gauge degrees of freedom as done in the previous
work \cite{harikumar}, we will keep the full set of the
constraints (\ref{pri-con}) and (\ref{sec-con}). In order for
studying the whole constraints efficiently, we further need to
modify the constraints (\ref{sec-con}) as
\begin{eqnarray}
\label{new-con}
\Lambda^\prime&\equiv&\partial^i\Omega_i+\Lambda\nonumber\\
&=&\partial^i\pi_i+A^0+\frac{1}{4m}\epsilon^{ijk}\partial_iB_{jk}\approx
0,\nonumber\\
\Lambda^\prime_i&\equiv&\partial^j\Omega_{ij}-\Lambda_i\nonumber\\
&=&\partial^j\pi_{ij}+B_{0i}-\frac{1}{2m}\epsilon_{ijk}
\partial^jA^k\approx 0,
\end{eqnarray}
which are equivalent to the original ones, $\Lambda$, $\Lambda_i$
on the constraint surface. Then, the set of these new constraints
makes the constraint algebra (\ref{pb}) much simpler and concise
as
\begin{eqnarray}
\label{con-algebra}
\{\pi_0, \Lambda'\}&=& -\delta(x-y),\nonumber\\
\{\pi_{0i}, \Lambda'_j \}&=& \delta_{ij}\delta(x-y),\nonumber\\
\{\Omega_i, \Omega_{jk} \}&=&
-\frac{1}{m}\epsilon_{ijk}\delta(x-y),
\end{eqnarray}
while the others identically vanish\footnote{From now on, we will
omit the prime symbols on the constraints $\Lambda$, $\Lambda_i$}.

As results, we have obtained the fully second-class constraints
(\ref{pri-con}) and (\ref{new-con}) for the first order Lagrangian
of the topologically massive theory. Notice that this new algebra
contains no derivatives unlikely in Eq. (\ref{pb}) as well as
vanishing brackets between the constraints $\Omega_i$,
$\Omega_{ij}$ and $\Lambda$, $\Lambda_i$. In fact, due to the
absence of the derivatives in the new Poisson brackets, one can
easily convert the second-class constraints (\ref{pri-con}) and
(\ref{new-con}) into corresponding first-class ones making use of
the BFT embedding technique, which we will explicitly study in the
next section. If we use the constraint algebra as it stands in Eq.
(\ref{pb}) containing the derivatives, BFT embedded constraints
and fields may have non-local expressions which would make the
quantization intractable.

On the other hand, making use of the definition of the Dirac
brackets as
\begin{eqnarray}
\label{def-db} \{A(x), B(y) \}_{DB} &=&\{A(x),B(y)\}_{PB} \nonumber\\
&-&\int dwdz \{A(x), \phi_\alpha (w)\}C^{\alpha\beta}(w,z)
\{\phi_\beta(z), B(y)\}, \nonumber
\end{eqnarray}
where the matrix $C^{\alpha\beta}$ is an inverse of
$\{\phi_\alpha(x),\phi_\beta(y)\}=C_{\alpha\beta}(x,y)$ along with
the constraints denoted by $\phi_\alpha=(\pi_0, \pi_{0i},\Lambda,
\Lambda_i, \Omega_i, \Omega_{ij})$, we have obtained the following
non-vanishing Dirac Brackets
\begin{eqnarray} \label{dirac-bra} \{A^0(x), A^i(y)\}_{D} &=&
\partial^i_x \delta(x-y),
\nonumber\\
 \{A^0(x), \pi_{ij}(y)\}_{D} &=& -\frac{1}{2m}\epsilon_{ijk}\partial^k_x
 \delta(x-y),
\nonumber\\
 \{A^i(x), \pi_j(y)\}_{D} &=& \frac{1}{2}\delta^i_j \delta(x-y),
\nonumber\\
 \{A^i(x), B^{jk}(y)\}_{D} &=& -m\epsilon^{ijk}\delta(x-y),
\nonumber\\
 \{\pi_i(x), B_{0j}(y)\}_{D} &=& \frac{1}{2m}\epsilon_{ijk}\partial^k_x
 \delta(x-y),
\nonumber\\
 \{\pi_i(x), \pi_{jk}(y)\}_{D} &=& \frac{1}{4m}\epsilon_{ijk} \delta(x-y),
\nonumber\\
 \{B^{0i}(x), B^{jk}(y)\}_{D} &=& (\delta^{ij}\partial^k_x - \delta^{ik}
 \partial^j_x) \delta(x-y),
\nonumber\\
 \{B^{ij}(x), \pi_{kl}(y)\}_{D} &=& \frac{1}{2}(\delta^i_k\delta^j_l-\delta^j_k\delta^i_l)
 \delta(x-y)
\end{eqnarray}
in order to compare with the results obtained from the BFT
embedding which automatically leads to the Dirac brackets at the
level of Poisson brackets in the extended phase space.

It seems appropriate to comment on the constraints $\Omega_i$, and
$\Omega_{ij}$. These constraints come from the topological term in
the Lagrangian (\ref{Lag}). In the scheme of Faddeev-Jackiw
quantization \cite{fj} which deals with only the dynamical degrees
of freedom, these could be eliminated from the start while
modifying other brackets of the fields. Harikumar and Sivakumar
\cite{harikumar} has worked in this direction. However, in the
present paper, we will keep these constraints with the others as
it stands and embed the whole constraints into much larger phase
space than their phase space, which we mean complete BFT
embedding. This will give a new type of WZ action with novel
symmetry as well as the St\"uckelberg action with the usual gauge
symmetry.

\section{Complete BFT Hamiltonian Embedding}
\setcounter{equation}{0}
\renewcommand{\theequation}{\arabic{section}.\arabic{equation}}

Since we know how to quantize first-class constraint system very
well while second-class system may have serious ordering and/or
non-local problems, it is preferred to deal with first-class
constraint system. The BFT embedding prescription makes in a
systematic way second-class constraint Hamiltonian system into
corresponding first-class one. In order for that purpose, we
introduce auxiliary fields having involutive relations in which
not only modified new constraints in the enlarged space are
strongly vanishing with each other but also they have vanishing
Poisson brackets, not the Dirac brackets, with physical quantities
such as Hamiltonian and fields themselves. Here, `physical' means
gauge invariant since resulting modified new quantities would be
first-class by construction. Practically, with the aid of
auxiliary fields $\Phi^\alpha$, one for each constraint,
satisfying with
\begin{equation}
\{\Phi^\alpha(x), \Phi^\beta(y)\}= \omega^{\alpha\beta}(x,y),
\end{equation}
while vanishing with the original fields, we construct the
involutive relations as
\begin{eqnarray}
\label{inv-const}
\{\tilde{\varphi}_\alpha(x), \tilde\varphi_\beta(y) \} &=& 0, \\
\label{inv-quant} \{\tilde{\varphi}_\alpha(x), \tilde{\cal F}(y)
\} &=& 0,
\end{eqnarray}
where the new constraints $\tilde{\varphi}_\alpha$, and physical
quantities $\tilde{\cal F}$ are given by
$\tilde{\varphi}_\alpha\sim\phi_\alpha
+\sum_{n=1}(\Phi^\alpha)^n$, and $\tilde{\cal F}\sim {\cal F}
+\sum_{n=1}(\Phi^\alpha)^n$, respectively, with appropriate
coefficients in front of $(\Phi^\alpha)^n$. In case we set the
auxiliary fields $\Phi^\alpha$ to be zero, those quantities are
reduced to the original ones, {\it i.e.},
$\tilde{\varphi}_\alpha\mid_{\Phi^\alpha=0}=\phi_\alpha$ (or,
$\tilde{\cal F}\mid_{\Phi^\alpha=0}={\cal F}$).

Now, let us explicitly solve the involutive relations, Eqs.
(\ref{inv-const}) and (\ref{inv-quant}), by making use of the
following Ansatz of
\begin{equation}
\label{ansatz} \varphi^{(1)}_\alpha(x)=\int dy~
X_{\alpha\beta}(x,y)\Phi^\beta(y).
\end{equation}
Inserting the Ansatz to the involutive relations between the
constraints (\ref{inv-const}), we obtain in the zeroth order of
the auxiliary fields $\Phi^\alpha$ as
\begin{eqnarray}
\label{sol-x} 0 &=&\label{reln} \{\varphi_\alpha,
\varphi_\beta\}_{\cal O}
+\{\varphi^{(1)}_\alpha, \varphi^{(1)}_\beta\}_{\Phi} \nonumber\\
&=& C_{\alpha\beta}(x,y)+\int dwdz~
X_{\alpha\gamma}(x,w)\omega^{\gamma\delta}(w,z)X_{\delta\beta}(z,y),
\end{eqnarray}
and in the first order of $\Phi^\alpha$ as
\begin{equation}
\label{first-order} \{\varphi_\alpha, \varphi^{(1)}_\beta\}_{\cal
O} +\{\varphi^{(1)}_\alpha, \varphi_\beta\}_{\cal O}
+\{\varphi^{(1)}_\alpha, \varphi^{(2)}_\beta\}_{\Phi}
+\{\varphi^{(2)}_\alpha, \varphi^{(1)}_\beta\}_{\Phi}=0,
\end{equation}
and so on. Here, the subscript $\cal O$(or, $\Phi$) denotes the
Poisson bracket with respect to the original variables $(q,p)$(or,
the auxiliary fields, $\Phi$). Then, by knowing the first order
correction $\varphi^{(1)}_\alpha$, we can obtain the next order
$\varphi^{(2)}_\alpha$ from Eq. (\ref{first-order}), and these
procedure continue iteratively until we determine involutive new
constraints completely.

Through the similar steps as above, we can also get the first
order correction for physical quantity as
\begin{equation}
\label{sol-phy} {\cal F}^{(1)}(x)=-\int dy dz dw ~\Phi^\alpha(x)
\omega_{\alpha\beta}(x,y)X^{\beta\gamma}(y,z)\{\varphi_\gamma(z),
A(w)\},
\end{equation}
in the zeroth order of the auxiliary fields $\Phi^\alpha$. This
correction has been obtained explicitly from the ansatz
(\ref{ansatz}) and the involutive relation (\ref{inv-quant}).
Also, higher order corrections ${\cal F}^{(n)}$ $(n\ge 2)$ can be
obtained iteratively from Eq. (\ref{inv-quant}) to find completely
involutive quantities out along with the modified new constraints.

It is appropriate to comment on the choice of
$\omega^{\alpha\beta}$ and $X_{\alpha\beta}$ which may be the
functions of the original variables. Although there are no known
criteria to best choice $\omega^{\alpha\beta}$ and
$X_{\alpha\beta}$, but final forms of the modified constraints and
the physical quantities depend on what the matrix elements of
$\omega^{\alpha\beta}$ and $X_{\alpha\beta}$ are. Nevertheless,
among different expressions of constraints and physical quantities
obtained from choosing any set of $\omega^{\alpha\beta}$ and
$X_{\alpha\beta}$, they are related to each other with canonical
transformations.

Now return to our model. By following the above BFT prescription,
we introduce auxiliary fields paired as $(\theta, \pi_\theta)$,
$(Q^i, P_i)$, and $(\Phi^i, \Phi^{jk})$:
\begin{eqnarray}
\Phi^\alpha=(\theta, Q^i, \pi_\theta, P^i, \Phi^i, \Phi^{ij}),
\nonumber
\end{eqnarray}
which correspond to the constraints $\phi_\alpha=(\pi_0, \pi_{0i},
\Lambda, \Lambda_i, \Omega_i, \Omega_{ij})$, respectively. Without
any loss of generality, let us for simplicity make a proper choice
for the auxiliary fields canonically conjugated as follows
\begin{equation}\label{omega}
\{\Phi^\alpha(x), \Phi^\beta(y) \}=\omega^{\alpha\beta}(x,y) =
\left(
\begin{array}{cccccc}
0 & 0 & 1 & 0 & 0 & 0\\
0 & 0 & 0 & \delta^{ij} & 0 & 0\\
-1 & 0 & 0 & 0 & 0 & 0 \\
0 & - \delta^{ij} & 0 & 0 & 0 & 0 \\
0 & 0 & 0 & 0 & 0 & \epsilon^{ijk} \\
0 & 0 & 0 & 0 & -\epsilon^{ijk}& 0
\end{array}\right)\delta(x-y).
\end{equation}
Here, note that the auxiliary fields $\Phi^i$, $\Phi^{ij}$ in the
last two columns is explicitly given by
\begin{eqnarray}
\label{sym-ext} \{\Phi^i(x), \Phi^{jk}(y)\}=
\epsilon^{ijk}\delta(x-y),
\end{eqnarray}
which choice makes it possible to embed the constraints originated
from the topological term. With the above choice of
$\omega^{\alpha\beta}$, we solve the Eq. (\ref{sol-x}) and find a
simple solution as
\begin{equation}
\label{x} X_{\alpha\beta}(x,y)=\left(
\begin{array}{cccccc}
1 & 0 & 0 & 0 & 0 & 0 \\
0 & \delta^{ij} & 0 & 0 &  0 & 0 \\
0 & 0 & 1 & 0 & 0 & 0 \\
0 & 0 & 0 & \delta^{ij}& 0 & 0 \\
0 & 0 & 0 & 0 & I_{3\times 3} & 0 \\
0 & 0 & 0 & 0 & 0 & \frac{1}{m}I_{3\times 3}
\end{array} \right)\delta(x-y),
\end{equation}
where $I_{3\times 3}$ denotes a $3\times 3$ identity matrix.

As results, we have obtained the strongly involutive new
constraints from Eq. (\ref{ansatz}) as
\begin{eqnarray}
\label{involutive-constraints} \widetilde{\pi}_0 &=& \pi_0
+\theta,~~~
\widetilde{\pi}_{0i} = \pi_{0i} + Q_i, \nonumber\\
\widetilde{\Lambda} &=& \Lambda + \pi_\theta, ~~~
\widetilde{\Lambda}_i = \Lambda_i + P_i, \nonumber\\
\widetilde{\Omega}_i &=& \Omega_i + \Phi_i, ~~~
\widetilde{\Omega}_{ij} = \Omega_{ij}+ \frac{1}{m}\Phi_{ij}
\end{eqnarray}
with no higher order contributions of the auxiliary fields due to
our proper choice of $\omega^{\alpha\beta}$ and $X_{\alpha\beta}$
while satisfying the involutive relation ({\ref{inv-const}}).
Moreover, from the inverse matrix $\omega_{\alpha\beta}$ and
$X^{\alpha\beta}$ and the solution of the physical fields
(\ref{sol-phy}), we have also gotten the strongly involutive
physical fields \cite{phy-field} as follows
\begin{eqnarray}
\label{physfield}
\widetilde{A}^0 &=& A^0 + \pi_\theta, \nonumber\\
\widetilde{A}^i &=& A^i + \partial^i\theta -
\frac{1}{2}\epsilon^{ijk}\Phi_{jk}, \nonumber \\
\widetilde{\pi}_0 &=& \pi_0 + \theta, \nonumber\\
\widetilde{\pi}_i &=& \pi_i - \frac{1}{2m}\epsilon_{ijk}\partial^j
Q^k + \frac{1}{2}\Phi_i, \nonumber \\
\widetilde{B}^{0i} &=& B^{0i}+P^i, \nonumber\\
\widetilde{B}^{ij} &=& B^{ij}-(\partial^iQ^j-\partial^jQ^i)+
m\epsilon^{ijk}\Phi_k, \nonumber\\
\widetilde{\pi}_{0i} &=& \pi_{0i}+Q_i, \nonumber\\
\widetilde{\pi}_{ij} &=& \pi_{ij}-\frac{1}{2m}\epsilon_{ijk}
\partial^k \theta + \frac{1}{2m}\Phi_{ij}
\end{eqnarray}
satisfying the involutive relation (\ref{inv-quant}). Note that
all the physical fields are terminated in the first order of the
auxiliary fields of the BFT embedding sequence and there are no
higher order contributions. One can easily checked that the
Poisson brackets of Eq. (\ref{physfield}) in the extended phase
space give exactly the same Dirac brackets in the original phase
space as given in Eq. (\ref{dirac-bra}) as they should be.

On the other hand, canonical Hamiltonian in the extended phase
space can be obtained either by solving the strongly involutive
relation (\ref{inv-quant}) in replacement of $\tilde{\cal F}$ with
a Hamiltonian function or by using the physical fields
(\ref{physfield}) in the canonical Hamiltonian (\ref{can-H})
written by the tilde fields. Since these methods of obtaining
canonical Hamiltonian are known to be equivalent to each other on
the constraint surface, which are weakly equivalent, we will
follow the latter approach. Then, the canonical Hamiltonian
(\ref{can-H}) is simply written in the extended phase space as
follows
\begin{eqnarray}
\label{ext-cH} \widetilde{\cal H}_c &=&
\frac{1}{4}(B_{ij}-Q_{ij})^2+\frac{1}{2}m\epsilon_{ijk}(B^{ij}-Q^{ij})\Phi^k
-\frac{1}{2}m^2\Phi_i\Phi^i \nonumber\\
&-&\frac{1}{2}(A_i+\partial_i\theta)^2+\frac{1}{2}\epsilon_{ijk}
(A^i+\partial^i\theta)\Phi^{jk}+\frac{1}{4}\Phi_{ij}\Phi^{ij}
\nonumber\\
&+&\frac{1}{2}(B_{0i}+P_i)^2-\frac{1}{m}\epsilon_{ijk}(B^{0i}+P^i)
\partial^jA^k-\frac{1}{m}(B_{0i}+P_i)\partial_j\Phi^{ij}
\nonumber\\
&-&\frac{1}{2}(A_0+\pi_\theta)^2-\frac{1}{2m}(A_0+\pi_\theta)\epsilon_{ijk}
\partial^iB^{jk}+(A_0+\pi_\theta)\partial_i\Phi^i,
\end{eqnarray}
where $Q_{ij}$ denote $Q_{ij}=\partial_i Q_j - \partial_j Q_i$. By
construction, this canonical Hamiltonian satisfies the strongly
involutive relations with the modified constraints
$\tilde{\varphi}_\alpha$, {\it i.e.},   $\{\tilde{\varphi}_\alpha,
\widetilde {\cal H}_c \}=0$. This ends the BFT embedding for the
topologically massive theory with the one- and two-form fields.

\section{Corresponding Lagrangian with new type of WZ term}
\setcounter{equation}{0}
\renewcommand{\theequation}{\arabic{section}.\arabic{equation}}

In this section, we find the corresponding Lagrangian of the
extended canonical Hamiltonian by making use of the phase space
path integral. For this purpose we modify further the canonical
Hamiltonian (\ref{ext-cH}) to an equivalent one on the constraint
surface which generates the Gauss' constraints naturally. The
equivalent Hamiltonian given by
\begin{equation}
\label{ext-mH} \widetilde{\cal H}'_c = \widetilde{\cal H}_c +
\pi_\theta \widetilde{\Lambda}+P_i\widetilde{\Lambda}_i
\end{equation}
yields the Gauss' constraints as
\begin{eqnarray}
\frac{d}{dt}\widetilde{\pi}_0 &=& \{\widetilde{\pi}_0,
\widetilde{\cal H}'_c \} = \widetilde{\Lambda}, \nonumber\\
\frac{d}{dt}\widetilde{\pi}_{0i} &=& \{\widetilde{\pi}_{0i},
\widetilde{\cal H}'_c \} = \widetilde{\Lambda}^i.
\end{eqnarray}

\subsection{Extended Gauge Symmetries}

Now, in order for obtaining the corresponding Lagrangian, we write
the generating functional of the topologically massive theory in
the fully extended phase space as
\begin{eqnarray}
\label{path-int} {\cal Z}&=&\int {\cal D}A^\mu {\cal D}\pi_\mu
{\cal D}B^{\mu\nu} {\cal D}\pi_{\mu\nu} {\cal D}\theta {\cal
D}\pi_\theta {\cal D}Q^i {\cal D} P_i {\cal D}\Phi^i {\cal
D}\Phi^{ij}\nonumber\\
&\times&\delta(\widetilde{\varphi}_\alpha)\delta(\Gamma_\beta)~
{\rm det}\mid\{\widetilde{\varphi}_\alpha, \Gamma_\beta \}\mid
e^{iS},
\end{eqnarray}
where
\begin{eqnarray}
S&=&\int d^4x ~\left[\pi_\mu\dot{A}^\mu+\pi_{0i}\dot{B}^{0i}
+\frac{1}{2}\pi_{ij}\dot{B}^{ij}+\pi_\theta\dot{\theta}+P_i\dot{Q}^i
+\frac{1}{2}\epsilon_{ijk}\Phi^i\dot{\Phi}^{jk}-\widetilde{\cal
H}'_c \right],\nonumber\\
\end{eqnarray}
and $\Gamma_\alpha$ are appropriate gauge fixing functions which
have non-vanishing Poisson brackets with the modified first-class
constraints $\tilde{\varphi}_\alpha$.

First, we can easily integrate the momenta variable, $\pi_0$,
$\pi_i$, $\pi_{0i}$, and $\pi_{ij}$ out along with the delta
functional, $\delta(\widetilde{\pi}_0)$,
$\delta(\widetilde{\Omega}_i)$, $\delta(\widetilde{\pi}_{0i})$,
and $\delta(\widetilde{\Omega}_{ij})$, respectively. Then, making
use of the Fourier transformations of the constraints
$\tilde{\Lambda}$, $\tilde{\Lambda}_i$ as
$\delta(\widetilde{\Lambda})=\int {\cal D}\xi \exp(-i\int
d^4x~\xi\widetilde{\Lambda})$, $\delta(\widetilde{\Lambda}_i)=\int
{\cal D}\chi^i \exp(-i\int d^4x~\chi^i\widetilde{\Lambda}_i)$, and
transforming  $A^0 \rightarrow A^0+\xi$, $B^{0i} \rightarrow
B^{0i}-\chi^i$, and after the Gaussian integrations over
$\pi_\theta$ and $P_i$ variables, we finally obtain the generating
functional as
\begin{equation}\label{generating-ftn}
{\cal Z}=\int {\cal D}A^\mu {\cal D}B^{\mu\nu} {\cal D}\theta
{\cal D}Q^i {\cal D}\Phi^i {\cal D}\Phi^{ij}{\cal D}Q^0
\delta(Q^0)\delta(\Gamma_\beta)
det\mid\{\widetilde{\varphi}_\alpha, \Gamma_\beta \}\mid e^{iS_T},
\end{equation}
where
\begin{eqnarray}\label{tot-act}
S_T & = & \int d^4x~\left( {\cal L}_{St} + {\cal
L}_{NWZ}\right) \\
\label{stuckelberg} {\cal L}_{St} &=&
\frac{1}{2}(A_\mu+\partial_\mu\theta)^2 -\frac{1}{4}
(B_{\mu\nu}-Q_{\mu\nu})^2 +
\frac{1}{2m}\epsilon_{\mu\nu\rho\sigma}
(B^{\mu\nu}-Q^{\mu\nu})\partial^\rho(A^\sigma+\partial^\sigma\theta)
\nonumber \\ \\
\label{nwz}{\cal L}_{NWZ} &=& \left[(\partial_i A_0-\partial_0
A_i)- \frac{m}{2}\epsilon_{ijk} (B^{jk}-Q^{jk})+
\frac{m^2}{2}\Phi_i\right]\Phi^i \nonumber\\
&-&\left[\frac{1}{2}\epsilon_{ijk}(A^i+\partial^i\theta)+\frac{1}{m}\partial_k
B_{0j}+\frac{1}{2m}\partial_0
B_{jk}+\frac{1}{4}\Phi_{ij}\right]\Phi^{jk} \nonumber\\
&+&\frac{1}{2}\epsilon_{ijk}\Phi^i\dot{\Phi}^{jk}.
\end{eqnarray}

Next, let us construct the gauge transformation generator $G$,
following Dirac's conjecture \cite{dirac}, for the embedded theory
in the standard way,
\begin{equation}
G=\int d^4x \sum_\alpha \epsilon^\alpha
\widetilde{\varphi}_\alpha,
\end{equation}
where $\widetilde{\varphi}_\alpha=(\widetilde{\pi}_0,
\widetilde{\Omega}_i, \widetilde{\Lambda}, \widetilde{\pi}_{0i},
\widetilde{\Omega}_{ij}, \widetilde{\Lambda}_i)$ are  the
first-class constraints in equation (\ref{involutive-constraints})
in order, and $\epsilon^\alpha = (\epsilon^{0}_{A},
\epsilon^{i}_{A}, \epsilon_{A}, \epsilon^{0i}_{B},
\epsilon^{ij}_{B}, \epsilon^{i}_{B})$ are, in general, functions
of phase space variables. The infinitesimal gauge transformation
for a function $F$ of phase space variables is then given by the
relation of $\delta F =\{F, G\}_D$, and leads to
\begin{eqnarray}
\label{ext-gt} \delta A^0&=&\epsilon^0_A,~~~~~~~~~~~~~~ \delta
B^{0i}=\epsilon^{0i}_B,\nonumber\\
\delta A^i&=&\epsilon^i_A-\partial^i\epsilon_A,~~~~~ \delta
B^{ij}=\partial^i\epsilon^j_B-\partial^j\epsilon^i_B+
\epsilon^{ij}_B-\epsilon^{ji}_B, \nonumber\\
\delta\theta&=&\epsilon_A,~~~~~~~~~~~~~~ \delta Q^i = \epsilon^i_B, \nonumber\\
\delta\Phi^i&=&\frac{1}{m}\epsilon^{ijk}\epsilon^B_{jk},~~~~~~
\delta\Phi^{ij}=-\epsilon^{ijk}\epsilon^A_k.
\end{eqnarray}

The above gauge transformation involving the gauge parameters is a
symmetry of the Hamiltonian, but not of the Lagrangian. The
generator $G$ of the most general local symmetry transformation of
a Lagrangian must satisfy the master equation \cite{rothe99}
\begin{equation}
\frac{\partial G}{\partial t}+\{G, H_T\}=0,
\end{equation}
which, together with (\ref{ext-gt}), implies the following
restrictions on the gauge parameters and on the Lagrangian
multipliers in the primary Hamiltonian:
\begin{eqnarray}\label{mul-rel}
\delta v^\beta &=&
\frac{d\epsilon^\beta}{dt}-\epsilon^P(V^\beta_P+v^\alpha
C^\beta_{\alpha_P}), \nonumber\\
0&=&\frac{d\epsilon^b}{dt}-\epsilon^P(V^b_P+v^\alpha
C^b_{\alpha_P}).
\end{eqnarray}
Here the superscripts $\alpha$, $\beta$, ($a$, $b$) denote the
primary (secondary) constraints, and $V^P_Q$, $C^P_{QR}$ are the
structure functions of the constrained Hamiltonian dynamics
defined by $\{H_c, \widetilde{\varphi}_P\}_D =
V^Q_P\widetilde{\varphi}_Q$, $\{\widetilde{\varphi}_P,
\widetilde{\varphi}_Q\}=C^R_{PQ}\widetilde{\varphi}_R$,
respectively. From (\ref{mul-rel}) we obtain
$\epsilon^0_A=-d\epsilon_A/dt$ and
$\epsilon^{0i}_B=d\epsilon^i_B/dt$. Thus, the gauge
transformations of $A^0$ and $B^{0i}$ in Eq. (\ref{ext-gt}) reduce
to
\begin{equation}
\delta A^0 = -\frac{d\epsilon_A}{dt}, ~~ \delta
B^{0i}=\frac{d\epsilon^i_B}{dt}.
\end{equation}
Redefining the gauge parameters \cite{henneaux85} as
$\bar{\epsilon}^i_B=\epsilon^i_B-\partial^i\int dt \epsilon^0_B$
along with the variation $\delta Q^0=\epsilon^0_B$, the final
gauge transformations\footnote{We have omitted the bar symbol in
the final rules of gauge transformations.} are nicely summarized
as
\begin{eqnarray}
\label{fin-ext-gt} \delta A^\mu &=&
-\partial^\mu\epsilon_A+\delta^\mu_j \epsilon^j_A,~~ \delta
B^{\mu\nu}=\partial^\mu\epsilon^\nu_B -
\partial^\nu\epsilon^\mu_B+(\epsilon^{kl}_B-\epsilon^{lk}_B)
\delta^\mu_k\delta^\nu_l, \nonumber\\
\delta\theta &=& \epsilon_A, ~~~~~~~~~~~~~~~~~\delta Q^\mu =
\epsilon^\mu_B,
\nonumber\\
\delta\Phi^i&=&\frac{1}{m}\epsilon^{ijk}\epsilon^B_{jk},~~~~~~~~~
\delta\Phi^{ij}=-\epsilon^{ijk}\epsilon^A_k.
\end{eqnarray}
Note here that the gauge parameter $\epsilon_A$ is related with
the topologically massive one-form fields $A^\mu$ and their WZ
field $\theta$ while the gauge parameters $\epsilon^\mu_B$ are
connected with the two-form fields $B^{\mu\nu}$ and their WZ
fields $Q^\mu$ resulting in the usual symmetry of the
St\"uckelberg Lagrangian. On the other hand, the gauge parameters
$\epsilon^i_A$, $\epsilon^{ij}_B$ are obtained from the embedded
topological constraints $\tilde{\Omega}_i$, $\tilde{\Omega}_{ij}$
with their WZ fields $\Phi^i$, $\Phi^{ij}$ resulting in the new
type of WZ Lagrangian. It is easy to check that the total action
(\ref{tot-act}) is exactly invariant under those gauge
transformation. As results, we have explicitly obtained the usual
gauge invariant St\"uckelberg Lagrangian ${\cal L}_{St}$ for the
massive one- and two-form fields, and a new type of the WZ
Lagrangian ${\cal L}_{NWZ}$ including the WZ fields, $\Phi^i$,
$\Phi^{ij}$.

It is appropriate to comment that in the generating functional
(\ref{generating-ftn}) there exists the delta functional of a
variable $Q^0$ which transforms as $\delta Q^0=\epsilon^0_B$ as
shown above. We have introduced this new field to make the final
Lagrangian manifestly covariant. Even without this $Q^0$ field, we
can show that the resulting Lagrangian successfully reproduces all
the BFT embedded constraint structure as in the section 3.
However, it fails to have manifest covariance. According to the
usage of the Hamiltonian formulation, the constraint structure of
the Lagrangian (\ref{Lag}) is called irreducible, in other words,
the constraints are linearly independent. On the other hand, the
constraint structure of the St\"uckelberg Lagrangian
(\ref{stuckelberg}) is reducible, {\it i.e.,} there is a redundant
relation among the constraints. In fact, we have introduced the
new variable $Q^0$ in order for keeping the manifest covariance,
while giving up the irreducible property between the constraints.

On the other hand, the new type of the WZ action is related to the
symplectic constraints, $\Omega_i$, $\Omega_{ij}$, of the
Lagrangian (\ref{Lag}), which are now converted into the
first-class constraints. This seemingly non-covariant form of the
Lagrangian ${\cal L}_{NWZ}$ comes from the introduction of the
auxiliary fields $\Phi^i$, $\Phi^{ij}$ as given by Eq.
(\ref{sym-ext}) where the distinction between the fields and the
momenta is useless. In other words, they look like another Dirac
brackets, and in order for embedding this symplectic structure
fully and getting a completely covariant action such as the
symplectic structure free theory, we may introduce infinite
numbers of auxiliary fields as discussed in Ref. \cite{kkp}.

\subsection{Various Gauge Fixings}

Now, let us consider various gauge fixings in the path integral
(\ref{path-int}). First, unitary gauge fixings: by fixing unitary
gauge means that all the auxiliary fields are set to be zero,
$\Phi_\alpha=0$, and it is easy task to check that we can recover
the original Lagrangian (\ref{Lag}) in the unitary gauge as
follows
\begin{equation}
{\cal Z}=\int {\cal D}A^\mu{\cal D}B^{\mu\nu} e^{iS},
\end{equation}
with the action $S$ in Eq. (\ref{Lag}). Furthermore, it is easy to
see that the integration over $B^{\mu\nu}$ yields the Proca
action, while the integration over $A^\mu$ Kalb-Ramond action.
Therefore, we have reconfirmed the well-known result that the
action (\ref{Lag}) is nothing but a master action leading to a
dual description, {\it i.e.}, they have a common origin.

Next, let us consider appropriate gauge fixings showing that the
topologically massive theory of first order with the one- and
two-form fields is equivalent to the $B\wedge F$ theory. From the
generating functional (\ref{path-int}), we first eliminate the
auxiliary fields $\Phi^i=0$, $\Phi^{ij}=0$ by fixing the unitary
gauge in part. Then, the resulting Lagrangian is simply the
St\"uckelberg ${\cal L}_{St}$. Next, we consider the following
gauge fixings:
\begin{eqnarray}
\label{gf}
\partial_iQ^i&=&0, \nonumber\\
\partial_i B^{0i} &=& 0, \nonumber\\
\chi_i &\equiv& A_i -\frac{1}{m}\epsilon_{ijk}\partial^jB^{0k}=0, \nonumber\\
\chi_{ij} &\equiv& B_{ij}-\frac{1}{m}\epsilon_{ijk}\partial^k
A^0=0.
\end{eqnarray}
The momenta, $\pi_i$, $\pi_{ij}$, $\pi_0$, $\pi_{0i}$, and $Q^0$
fields are integrated out along with the constraints,
$\delta(\widetilde{\Omega}_i)$, $\delta(\widetilde{\Omega}_{ij})$,
$\delta(\widetilde{\pi}_0)$, $\delta(\widetilde{\pi}_{0i})$, and
$\delta(Q^0)$, respectively. After the momenta $\pi_\theta$,
$P_i$, integration along with the constraints
$\widetilde{\Lambda}$, $\widetilde{\Lambda}_i$, we obtain the
following intermediate generating functional
\begin{equation}
{\cal Z}=\int {\cal D}A^\mu{\cal D}B^{\mu\nu}{\cal D}\theta {\cal
D}Q^i \delta(\chi_i)\delta(\chi_{ij})~ {\rm
det}\mid\{\widetilde{\varphi}_\alpha, \Gamma_\beta \}\mid e^{iS},
\end{equation}
where the action
\begin{equation}
S=\int d^4x~\left[\frac{1}{2m}\epsilon_{ijk}\dot{A}^iB^{jk} -
\frac{1}{2m}\epsilon_{ijk}\partial^iB^{jk}\dot{\theta} +
\frac{1}{m}\epsilon_{ijk}\dot{Q}^i\partial^jA^k - \widetilde{H}'_c
\right],
\end{equation}
and
\begin{equation}
\widetilde{H}'_c = \frac{1}{4}(B_{ij}-Q_{ij})^2-\frac{1}{2}(A_i+
\partial_i\theta)^2+\frac{1}{4m^2}F_{ij}F^{ij}-\frac{1}{2\cdot3!m^2}
H_{ijk}H^{ijk}.
\end{equation}
Here, we have used identities
\begin{eqnarray}
&&\epsilon_{ijk}\epsilon^{lmn}\partial^iB^{jk}\partial_lB_{mn}=
-\frac{2}{3}H_{ijk}H^{ijk}, \nonumber\\
&&\epsilon_{ijk}\epsilon^{ilm}
\partial^jA^k\partial_lA_m= \frac{1}{2}F_{ij}F^{ij}, \nonumber
\end{eqnarray}
with $\partial^iA_i=0$ which comes from the gauge fixing condition
$\partial^i\chi_i=0$. We have also denoted $H_{ijk}=\partial_i
B_{jk}+\partial_j B_{ki}+\partial_k B_{ij}$ and $F_{ij}=\partial_i
A_j -\partial_j A_i$. The remaining $\theta$ and $Q^i$
integrations yields the following action
\begin{eqnarray}
\label{int-int-Lag} S&=&\int
d^4x~\left[\frac{1}{2m}\epsilon_{ijk}\dot{A}^iB^{jk}-
\frac{1}{4}B_{ij}B^{ij}+\frac{1}{2}A_iA^i-\frac{1}{4m^2}F_{ij}F^{ij}
+\frac{1}{2\cdot3!m^2}H_{ijk}H^{ijk}\right. \nonumber\\
&&~~~~~~~\left.
-\frac{1}{8m^2}\epsilon_{ijk}\epsilon^{ilm}\partial^0B^{jk}\partial_0B_{lm}
-\frac{1}{2m^2}\partial_0A_i\partial^0A^i\right].
\end{eqnarray}
Now, making use of the gauge conditions (\ref{gf}) and identities
\begin{eqnarray}
&&-\frac{1}{8m^2}\epsilon_{ijk}\epsilon^{ilm}\partial^0B^{jk}\partial_0B_{lm}
= \frac{1}{4m^2}H_{0ij}H^{0ij}-\frac{1}{2m^2}B_{0i}\nabla^2B^{0i},
\nonumber \\
&&-\frac{1}{2m^2}\partial_0A_i\partial^0A^i=-\frac{1}{2m^2}F_{0i}F^{0i}
+ \frac{1}{2m^2}A_0\nabla^2A^0, \nonumber
\end{eqnarray}
and the equations also obtained from the gauge conditions
(\ref{gf})
\begin{eqnarray}
&&\nabla^2A^0=\frac{m}{2}\epsilon_{ijk}\partial^iB^{jk}, \nonumber\\
&& \nabla^2B^{0i}=-m\epsilon^{ijk}\partial^jA^k, \nonumber
\end{eqnarray}
we can manipulate the last two terms in the action
(\ref{int-int-Lag}) with the others to obtain the well-known
$B\wedge F$ theory of
\begin{eqnarray}
{\cal Z} &=& \int {\cal D}A^\mu{\cal D}B^{\mu\nu}{\rm
det}\mid\{\widetilde{\varphi}_\alpha, \Gamma_\beta \}\mid
e^{iS}\nonumber\\
S &=& \int
d^4x~\left[-\frac{1}{4}F_{\mu\nu}F^{\mu\nu}+\frac{1}{2\cdot
3!}H_{\mu\nu\rho}H^{\mu\nu\rho}+\frac{m}{4}\epsilon_{\mu\nu\rho\sigma}
B^{\mu\nu}F^{\rho\sigma} \right],\nonumber\\
\end{eqnarray}
where the fields $A^\mu$, $B^{\mu\nu}$ are scaled as $mA^\mu$,
$mB^{\mu\nu}$, respectively.

As a result, we have explicitly shown that the equivalence of the
topologically massive theory of first order with the one- and
two-form fields and the $B\wedge F$ theory on the level of path
integral.

\section{Revisit the Gauging of Topologically Massive Theory}
\setcounter{equation}{0}
\renewcommand{\theequation}{\arabic{section}.\arabic{equation}}

In this section, we will explicitly show the equivalence of gauged
massive theory of first order and the St\"uckelberg Lagrangian,
${\cal L}_{St}$, in which the Lagrangian is obtained from the BFT
embedding.

By gauging the fields $A^\mu$, $B^{\mu\nu}$ means the following
transformations
\begin{equation}
A^\mu \rightarrow A^\mu+\partial^\mu\theta,
~~~B^{\mu\nu}\rightarrow B^{\mu\nu}-Q^{\mu\nu},
\end{equation}
where $Q^{\mu\nu}=\partial^\mu Q^\nu-\partial^\nu Q^\mu$. Thus,
the gauged massive theory of first order Lagrangian (\ref{Lag}) is
described by
\begin{equation}
\label{gauged-Lg} {\cal L}_G =
-\frac{1}{4}(B_{\mu\nu}-Q_{\mu\nu})^2+\frac{1}{2}
(A_\mu+\partial_\mu\theta)^2
+\frac{1}{2m}\epsilon_{\mu\nu\rho\sigma}(B^{\mu\nu}-Q^{\mu\nu})
\partial^\rho(A^\sigma+\partial^\sigma\theta),
\end{equation}
which is invariant under the gauge transformations of
\begin{equation}
\delta A^\mu = \partial^\mu\epsilon,~~ \delta\theta=-\epsilon,~~
\delta
B^{\mu\nu}=\partial^\mu\epsilon^\nu-\partial^\nu\epsilon^\mu,
~~\delta Q^\mu=\epsilon^\mu.
\end{equation}
Here, we have redefined the gauge parameters as $\epsilon_A=
-\epsilon$, $\epsilon^\mu_B=\epsilon^\mu$ in which expression is
usually seen in the literature and take $\epsilon^i_A
=\epsilon^{ij}_B=0$ in the extended gauge symmetries
(\ref{fin-ext-gt}) in order to focus mainly on the gauging
technique. Note that as you already know through the note in the
previous sections by taking $\epsilon^\mu=\partial^\mu\lambda$,
$\delta B^{\mu\nu}$ have vanished clearly indicating that the
constraints are not all independent, {\it i.e.,} reducible.

In order for comparing this action with the St\"uckelberg
Lagrangian ${\cal L}_{St}$ in view of constraint structure, we do
partial integration of the terms $G_0\dot\theta$,
$H_{0i}\dot{Q}^i$ to $\-\dot{G}_0\theta$, $-\dot{H}_{0i}Q^i$,
respectively, in the gauged Lagrangian (\ref{gauged-Lg}). Then,
the canonical momenta are obtained as
\begin{eqnarray}
\pi_0 &=& -\theta,
~~~\pi_i=\frac{1}{4m}\epsilon_{ijk}B^{jk},\nonumber\\
\pi_{0i}&=&-Q^i,~~~\pi_{ij}=-\frac{1}{2m}\epsilon_{ijk}A^k,\nonumber\\
P_i&=&Q_{i0},~~~P_0=0,~~~~~\pi_\theta=\dot\theta,
\end{eqnarray}
from which we have the primary constraints as
\begin{eqnarray}
&& \Sigma_0 \equiv \pi_0+\theta \approx 0, \nonumber\\
&& \Sigma_i \equiv \pi_{0i}+Q^i\approx 0,\nonumber\\
&& \Omega_i \equiv \pi_i-\frac{1}{4m}\epsilon_{ijk}B^{jk}\approx 0,\nonumber\\
&& \Omega_{ij} \equiv \pi_{ij}+\frac{1}{2m}\epsilon_{ijk}A^k\approx 0,\nonumber\\
&& P_0 \approx 0.
\end{eqnarray}
We also get the canonical Hamiltonian as
\begin{eqnarray}
{\cal H}_c &=&
\frac{1}{4}(B_{ij}-Q_{ij})^2-\frac{1}{2}(A_i+\partial_i\theta)^2
+\frac{1}{2}B_{0i}B^{0i}-\frac{1}{2}P_iP^i-\frac{1}{2}A_0A^0
 \nonumber\\
&+& \frac{1}{2}\pi^2_\theta+ (P_i+B_{0i})\partial^iQ^0-\frac{1}{m}
\epsilon_{ijk}B^{0i}\partial^jA^k-\frac{1}{2m}A^0
\epsilon_{ijk}\partial^iB^{jk}.
\end{eqnarray}
From the time stability conditions of the primary constraints with
the canonical Hamiltonian, there are additional secondary
constraints as follows
\begin{eqnarray}
\Delta &\equiv& \frac{d}{dt}P_0 = \partial^i(P_i+B_{0i})\approx 0,
\nonumber\\
\Lambda &\equiv& \frac{d}{dt}\Sigma_0 = A_0 + \frac{1}{2m}
\epsilon_{ijk}\partial^iB^{jk}+\pi_\theta \approx 0, \nonumber\\
\Lambda_i &\equiv& \frac{d}{dt}\Sigma_i = -B_{0i}+
\frac{1}{m}\epsilon_{ijk}\partial^jA^k-P_i \approx 0.
\end{eqnarray}
It seems appropriate to comment on the linear independence of the
constraints. The combination of the above constraints $\Delta$,
$\Lambda_i$ vanishes {\it i.e.,} $\Delta+\partial^i \Lambda_i=0$,
which indicates again that the whole constraints are not linearly
independent, {\it i.e.,} reducible.

Finally, making use of the above constraints, we can obtain the
following Hamiltonian
\begin{eqnarray}
\label{ext-cH-revisit} {\cal H}_c &=& \frac{1}{4}(B_{ij}-Q_{ij})^2
-\frac{1}{2}(A_i+\partial_i\theta)^2 +\frac{1}{2}(B_{0i}+P_i)^2
-\frac{1}{m}\epsilon_{ijk}(B^{0i}+P^i)
\partial^jA^k\nonumber\\
&-&\frac{1}{2}(A_0+\pi_\theta)^2
-\frac{1}{2m}(A_0+\pi_\theta)\epsilon_{ijk}\partial^iB^{jk}
+\pi_\theta\Omega+(P^i+\partial^iQ^0)\Omega_i \nonumber\\
&\approx& \widetilde{\cal H}'_c,
\end{eqnarray}
where $\widetilde{\cal H}'_c$ is given in Eq. (\ref{ext-cH}). This
shows the equivalence of the gauged Lagrangian (\ref{gauged-Lg})
and the St\"uckelberg Lagrangian (\ref{stuckelberg}) on the
constraint surface.

\section{Conclusion}

In this paper, we have applied the complete BFT method to the
massive theory with one- and two-form fields. We have newly
analyzed the full set of constraint structure of the model having
no derivatives in the Poisson brackets, which is much simpler than
that of the previous work \cite{harikumar} and thus makes it
possible to apply the BFT embedding further. Then, we have
explicitly carried out the complete BFT embedding of the theory
including the gauge symmetry breaking terms and the topological
term, which was not gauge invariant. Exploiting the complete BFT
embedding we have obtained the gauge invariant Lagrangian
corresponding to the first class Hamiltonian, and by identifying
the auxiliary fields with the St\"uckelberg vector fields and new
type of WZ fields, we have shown simultaneously the St\"uckelberg
Lagrangian related to the explicit gauge breaking mass term and
the new type of WZ action to topological term having novel
symmetry. Furthermore, by analyzing the gauged version of the
theory we have also shown that the usual St\"uckelberg Lagrangian
is exactly equivalent to the BFT embedded one on the constraint
space after taking the unitary gauge $\Phi^i=0$, $\Phi^{ij}=0$.

\vskip 1.0cm The work of YWK was supported by the Korea Research
Foundation, Grant No. KRF-2002-075-C00007.

\end{document}